\documentclass[reprint,superscriptaddress,amsmath,amssymb]{revtex4-1}
\pdfoutput=1
\usepackage{float}
\usepackage{xcolor}
\usepackage{graphicx}
\usepackage{dcolumn}
\usepackage{bm}
\usepackage[colorlinks=true, linkcolor=blue, citecolor=blue, urlcolor=blue]{hyperref}
\newcommand{\ZIB}{Zuse Institute Berlin, Takustra{\ss}e 7, 14195 Berlin, Germany}
\newcommand{\JCM}{JCMwave GmbH, Bolivarallee 22, 14050 Berlin, Germany}

\begin{document}
	
\title{Modal analysis for nanoplasmonics with nonlocal material properties}

\author{Felix Binkowski}
\affiliation{\ZIB}
\author{Lin Zschiedrich}
\affiliation{\JCM}
\author{Martin Hammerschmidt}
\affiliation{\JCM}
\author{Sven Burger}
\affiliation{\ZIB}
\affiliation{\JCM}

\begin{abstract}
Plasmonic devices with feature sizes of a few nanometers exhibit
effects which can be described by the nonlocal hydrodynamic Drude model.
We demonstrate how to exploit contour integral methods for computing eigenfrequencies
and resonant states of such systems. 
We propose an approach for deriving the modal expansion of relevant physical observables.
We use the methods to perform a modal analysis for a metal nanowire.
All complex eigenfrequencies in a large frequency range and the corresponding
resonant states are computed. 
We identify those resonant states which are relevant for the
extinction cross section of the nanowire. 

\end{abstract}
\maketitle

\section{Introduction} \label{sec_i}  
Nanofabrication technologies
allow for a rapid progress in engineering nano-optical devices~\cite{Lindquist_2012}.
Plasmonic resonances are the center of attention for many topical applications
exploring new regimes of physics.
Examples comprise the demonstration of plasmonic lasers~\cite{Oulton_2009},
tailoring light emission of nanoantennas~\cite{Curto_Science_2010,Giannini_PlasmonNanoantennas_2011},
probing single molecules and nanoparticles by Raman scattering~\cite{Nie_Science_1997},
plasmonic photochemistry~\cite{Zhang_ChemRev_2018},
and quantum emitters interacting with metal nanoresonators~\cite{Chikkaraddy_2016}.

An adequate description of material dispersion plays
an important role for the investigation of light-matter interaction in
plasmonic structures~\cite{Murray_2007}.
In many cases, the material dispersion can be described by the Drude-Lorentz model
or by a rational function fit to measured material data~\cite{Sehmi_2017,Garcia-Vergara_2017}.
Such models are based on spatially local interactions between the light and the
free electron gas of the plasmonic scatterers~\cite{Johnson_1972}.
When the scatterers are at the size of a few nanometers,
nonlocal material models are required~\cite{Raza_2015}.
These models lead to additional resonances of the electromagnetic field with sub-nm wavelengths.
Recently, surface plasmon resonance blueshifts
have been observed in metal nanoparticles~\cite{Scholl_2012,Raza_Blueshift_2013}
which could be explained~\cite{Christensen_2014}
using the nonlocal hydrodynamic Drude model (HDM)~\cite{Boardman_1982}.
This model assumes that the motion of the electron gas behaves
as a hydrodynamic flow and allows for the investigation of nonlocal physical
effects~\cite{Ruppin_2001,Palomba_2008,Abajo_2008,Intravaia_2015,Toscano_2015,Schnitzer_ProcRoyal_2016,Moeferdt_2018}.

For the study of physical phenomena in nanoplasmonic systems, 
a deeper understanding of the effects based on the HDM is required.
A modal description is the most instructive
approach~\cite{Kristensen_ModeVol_2014,Lalanne_QNMReview_2018}.
In the case of local material models, numerically computed resonant states
of plasmonic systems have been successfully used to derive modal
expansions~\cite{Sauvan_QNMexpansionPurcell_2013,Hughes_PRA_2017,Zschiedrich_PRA_2018,Yan_PRB_2018}.
However, in the case of the HDM, a coupled system of equations
has to be solved~\cite{McMahon_PRL_2009,Hiremath_Hydro2D_2012,Toscano_2012,Hughes_Optica_2017}.
To the best of our knowledge, for this system, 
the computations of eigenfrequencies in a large frequency range
with corresponding resonant states and modal expansions
have not yet been reported.

In this work, we investigate plasmonic resonances based on the HDM.
We present a contour-integral-based framework for a modal analysis.
Typical physical observables are sesquilinear forms which involve a complex 
conjugation of the solution fields.
We propose a general approach for the computation
of modal sesquilinear quantities. 
The framework is applied to calculate the eigenfrequencies and corresponding resonant states of
a metal nanowire.  Furthermore, the modal extinction cross section of the nanowire
illuminated by plane waves is computed. This allows to classify the resonant states
of the nanowire into states which couple to the light sources and into states
which have no contribution to the extinction cross section. 

This work is structured as follows.
Section~\ref{sec_ii} introduces a coupled system of equations describing the HDM
and summarizes numerical methods for modal analysis.
In the subsection {\it Modal expansion of sesquilinear quantities},
we extend the framework of the Riesz projection expansion (RPE)~\cite{Zschiedrich_PRA_2018}
in order to obtain modal expansions of physical observables, such as
the extinction cross section.
Section~\ref{sec_iii} applies these methods for an investigation of
the resonances of a metal nanowire.
Section~\ref{sec_iv} concludes the study.

\section{Plasmonic resonances based on the hydrodynamic Drude model}\label{sec_ii}  
The HDM is based on the interaction of
a nonlocal polarization current and its resulting electric field.
In the frequency domain and for nonmagnetic materials, this
is described by the coupled system of equations,
\begin{align}
	\nabla \times &\mu_0^{-1} \nabla  \times \mathbf{E}(\mathbf{r},\omega)
	- \omega^2\epsilon_\mathrm{loc}(\mathbf{r},\omega) \mathbf{E}(\mathbf{r},\omega) \nonumber\\
	&= i \omega \mathbf{J}_{\mathrm{hd}}(\mathbf{r},\omega) + 
	i \omega \mathbf{J}(\mathbf{r},\omega), \label{coupled_system1} \\
	\beta^2 \nabla &\left( \nabla \cdot \mathbf{J}_{\mathrm{hd}}(\mathbf{r},\omega) \right)
	+\omega\left( \omega+i \gamma \right) \mathbf{J}_{\mathrm{hd}}(\mathbf{r},\omega) \nonumber \\
	& =	i \omega \omega_\mathrm{p}^2 \epsilon_0 \mathbf{E}(\mathbf{r},\omega), \label{coupled_system2}
\end{align}
for the electric field $\mathbf{E}(\mathbf{r},\omega)$ and the nonlocal hydrodynamic
current density $\mathbf{J}_{\mathrm{hd}}(\mathbf{r},\omega)$, where
$\mathbf{J}(\mathbf{r},\omega)$ is a given impressed current density,
$\omega$ is the frequency, $\epsilon_\mathrm{loc}(\mathbf{r},\omega)$
is the permittivity resulting from the local material response,
$\epsilon_0$ is the 
vacuum permittivity, and $\mu_0$ is the vacuum permeability.
The damping constant $\gamma$ and the plasma frequency $\omega_\mathrm{p}$
correspond to the local Drude model 
$\epsilon_\mathrm{d}(\omega) = \epsilon_0(\epsilon_\infty
 - \omega_\mathrm{p}^2/(\omega^2 +i \gamma \omega))$,
where $\epsilon_\infty$ is the relative permittivity at infinity.
The factor $\beta = \sqrt{3/5}\,v_\mathrm{F}$ relates to the Fermi velocity $v_\mathrm{F}$~\cite{Boardman_1982}.

\begin{figure}[]
	{\includegraphics[width=0.45\textwidth]{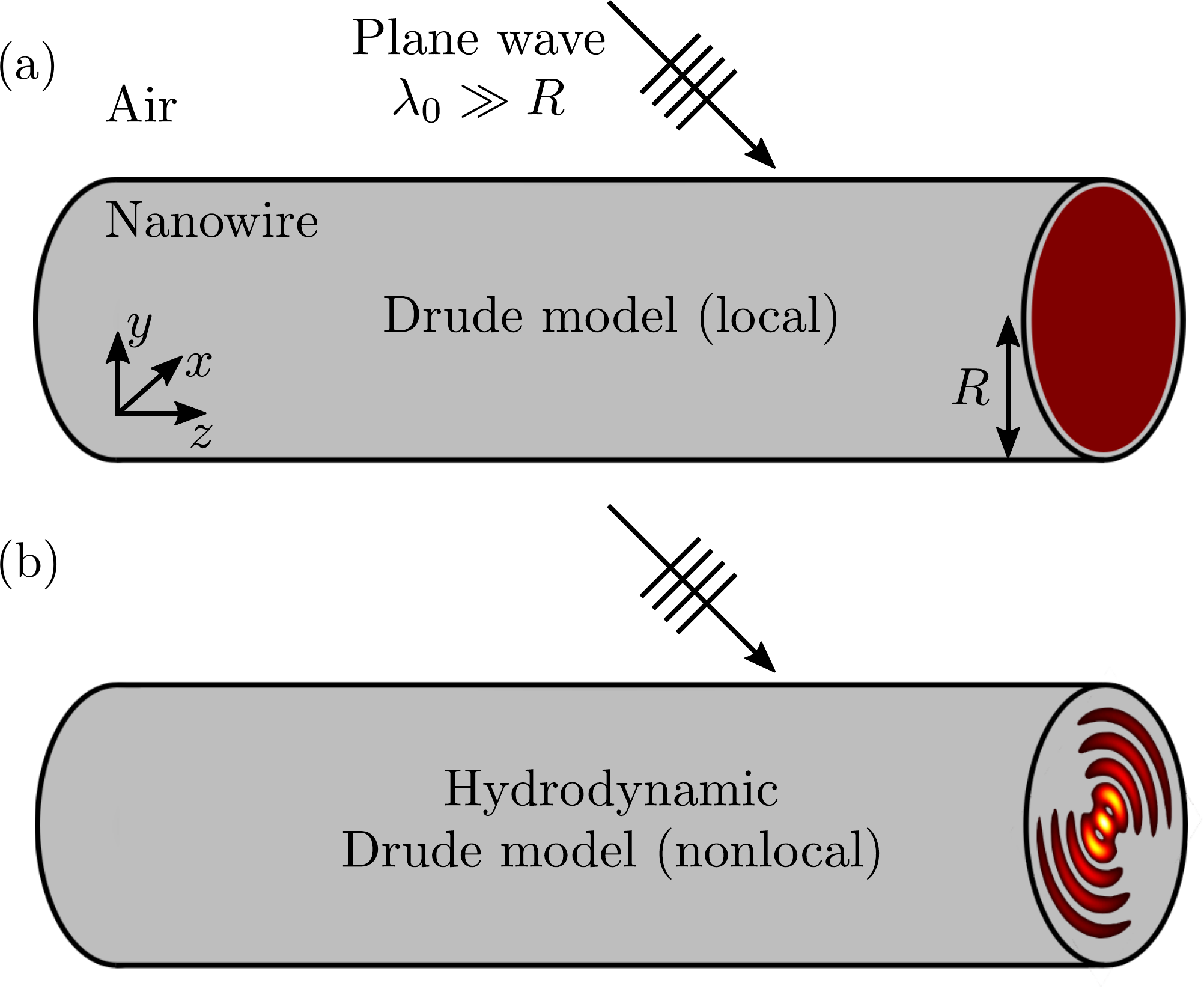}}
	\caption{Schematics of a metal nanowire illuminated by a plane wave of wavelength $\lambda_0$.
		Electric field intensity sketched on a cut through the nanowire.
		(a) Nearly constant electric field intensity in case of the local Drude model.
		(b) Radially oscillating field pattern in case of the nonlocal hydrodynamic Drude model.} 
	\label{fig:fig01}
\end{figure}

The nonlocal material response is caused by $\mathbf{J}_{\mathrm{hd}}(\mathbf{r},\omega)$,
which affects the permittivity function for the free electron gas.
If $\beta \rightarrow 0$, then the coupled system simplifies to Maxwell's equations
for the local Drude model. For an illustration of the effect of the HDM,
a nanowire excited by a plane wave is sketched in Fig.~\ref{fig:fig01}.
While, for the local Drude model, the electric field intensity inside of the nanowire
is nearly constant, the electric field pattern is radially
oscillating considering the HDM
[see~Figs.~\ref{fig:fig01}\textcolor{blue}{(a)} and \ref{fig:fig01}\textcolor{blue}{(b)}, respectively].
The reader is referred to \cite{Hiremath_Hydro2D_2012,Toscano_2012}
for a detailed derivation of Eqs.~\eqref{coupled_system1}
and \eqref{coupled_system2} including the applied assumptions and approximations. 

Physical scattering solutions $\mathbf{E}(\mathbf{r},\omega_0)$ and $\mathbf{J}_{\mathrm{hd}}(\mathbf{r},\omega_0)$
of the coupled system can be obtained for real frequencies $\omega_0 \in \mathbb{R}$.
The eigenfrequencies are defined as the complex resonance poles
$\tilde{\omega}_k \in \mathbb{C}$ of the analytical continuation of
$\mathbf{E}(\mathbf{r},\omega_0)$ and $\mathbf{J}_{\mathrm{hd}}(\mathbf{r},\omega_0)$
into the complex plane yielding $\mathbf{E}(\mathbf{r},\omega)$
and $\mathbf{J}_{\mathrm{hd}}(\mathbf{r},\omega)$, where $\omega \in \mathbb{C}$~\cite{Zschiedrich_PRA_2018}.
The resonant states, also called eigenmodes, of the coupled system
correspond to these eigenfrequencies.

\subsection{Numerical methods for modal analysis}
The contour integral method \textsc{Beyn's Algorithm}~\cite{Beyn_LAAppl_2012}
is applied to numerically solve the nonlinear
eigenproblem~\cite{Guettel_NLEVP_2017}
corresponding to the coupled system given by Eqs.~\eqref{coupled_system1}
and \eqref{coupled_system2}.
Contour integral methods for such problems require the definition of an integration path
in the complex frequency plane which encloses the eigenfrequencies corresponding
to the eigenmodes of interest. The numerical integration along this contour
projects vector fields onto the space spanned by these eigenmodes.
In this way, an approximate eigenspace is constructed.
Then, e.g., the methods proposed
in~\cite{Asakura_JSIAM_2009,Beyn_LAAppl_2012} apply
a singular-value decomposition (SVD) to this approximate
eigenspace and solve a linear eigenproblem of small dimension.
The approach presented in~\cite{Gavin_JCompPhy_2018} applies the
Rayleigh-Ritz method to the approximate eigenspace and solves a nonlinear
eigenproblem of small dimension.
The common property of these methods is that they
essentially require the solution of scattering problems for the integration points on
the chosen contour.
This is in contrast to standard approaches for solving nonlinear eigenproblems,
such as the Arnoldi method,
which are based on linearization of the nonprojected problems
using auxiliary fields~\cite{Saad_Book_NumMeth_Eig_2011,Brule_AuxFields_2016}.

For the modal expansion of scattering problems,
an unconjugated scalar product
can be used~\cite{Yan_PRB_2018}.
In this context, it is an open problem how to deal with the expansion of nonholomorphic
quantities, e.g., the extinction cross section.
The contour-integral-based RPE~\cite{Zschiedrich_PRA_2018}
allows one to perform a modal expansion without a scalar product.
A solution $\mathbf{E}(\mathbf{r},\omega_0)$
to the coupled system given by Eqs.~\eqref{coupled_system1}
and \eqref{coupled_system2} can be expanded into a weighted sum of
eigenmodes yielding the coupling of the modes to specific sources $\mathbf{J}(\mathbf{r},\omega_0)$
with $\omega_0 \in \mathbb{R}$.
Cauchy's integral formula,
\begin{align}
	\mathbf{E}(\mathbf{r},\omega_0) =
	\frac{1}{2 \pi i} \oint \limits_{C_0} 
	\frac{\mathbf{E}(\mathbf{r},\omega)}{\omega-\omega_0} \text{ d}\omega,
	\nonumber
\end{align}
is exploited, where $\mathbf{E}(\mathbf{r},\omega)$, $\omega \in \mathbb{C}$, is the analytical continuation of
$\mathbf{E}(\mathbf{r},\omega_0)$ into the complex plane and
$C_0$ is a closed integration path around $\omega_0$ so that $\mathbf{E}(\mathbf{r},\omega)$ 
is holomorphic inside of $C_0$. Deforming the integration path and applying Cauchy's residue theorem yield
\begin{align}
	\mathbf{E}(\mathbf{r},\omega_0)
	=  &- \frac{1}{2 \pi i} \oint\limits_{{\tilde{C}}_1}
	\frac{\mathbf{E}(\mathbf{r},\omega)}{\omega-\omega_0} \text{ d}\omega - 
	\dots - \frac{1}{2 \pi i} \oint\limits_{{\tilde{C}}_K}
	\frac{\mathbf{E}(\mathbf{r},\omega)}{\omega-\omega_0} \text{ d}\omega \nonumber \\
	&+ \frac{1}{2 \pi i}	\oint\limits_{C_{\text{nr}}}
	\frac{\mathbf{E}(\mathbf{r},\omega)}{\omega-\omega_0} \text{ d}\omega, \nonumber
\end{align}
where $\tilde{C}_1,\dots,\tilde{C}_K$ are contours around the eigenfrequencies $\tilde{\omega}_1,\dots,\tilde{\omega}_K$
and $C_{\text{nr}}$ is a contour including $\omega_0$, the eigenfrequencies
$\tilde{\omega}_1,\dots,\tilde{\omega}_K$, and no additional eigenfrequencies.
The Riesz projections,
\begin{align}
	{\tilde{\mathbf{E}}}_k (\mathbf{r},\omega_0) = -\frac{1}{2 \pi i}
	\oint\limits_{{\tilde{C}}_k} \frac{\mathbf{E}(\mathbf{r},\omega)}{\omega-\omega_0}
	\text{ d}\omega, \nonumber
\end{align}
corresponding to $\tilde{\omega}_k$ describe
the coupling of the eigenmodes to the considered source field. The field,
\begin{align}
	{\mathbf{E}}_{\text{nr}} (\mathbf{r},\omega_0) =  \frac{1}{2 \pi i}
	\oint\limits_{C_{\text{nr}}}
	\frac{\mathbf{E}(\mathbf{r},\omega)}{\omega-\omega_0} \text{ d}\omega,
	\nonumber
\end{align}
contains nonresonant components as well as components corresponding to
eigenfrequencies outside of the contour $C_{\text{nr}}$.
For this modal expansion approach, 
instead of projecting random vectors as for \textsc{Beyn's Algorithm},
the numerical integration is performed by solving the coupled system
using physical source fields at the integration points.

Equations~\eqref{coupled_system1} and \eqref{coupled_system2}
are spatially discretized with the finite element method (FEM)~\cite{Monk_2003,Weiser_FEM_2016}.
The FEM solver \textsc{JCMsuite} is used to solve scattering
problems. Perfectly matched layers (PMLs)
are applied to realize outgoing radiation conditions~\cite{Berenger_1994}.
High order polynomial ansatz functions and mesh refinements are used to reach a sufficient
numerical accuracy~\cite{Lalanne_QNM_Benchmark_2018}.
We write 
\begin{align}
	T({\omega}) v = f(\omega),
 \nonumber
\end{align}
for the coupled system given by Eqs.~\eqref{coupled_system1}
and \eqref{coupled_system2}, where $T({\omega}) \in \mathbb{C}^{n\times n}$ is the system matrix
resulting from the FEM discretization and $v\in\mathbb{C}^{n}$ is the vector corresponding to
$\mathbf{E}(\mathbf{r},\omega)$ and $\mathbf{J}_{\mathrm{hd}}(\mathbf{r},\omega)$.
The dimension $n$ results from the spatial mesh and from the degrees of the
polynomial ansatz functions of the FEM discretization.
The right-hand side $f(\omega)$ corresponds to the impressed current
density $\mathbf{J}(\mathbf{r},\omega)$ and incoming source fields. In this notation,
$T({\tilde{\omega}_k}) \tilde{v}_k = 0$ holds
for an eigenfrequency $\tilde{\omega}_k$ and an eigenmode $\tilde{v}_k$.
Solving $T(\omega)v = f(\omega)$ with $f(\omega)\neq 0$
corresponds to solving a scattering problem. 

\subsection{Modal expansion of sesquilinear quantities}
Typical physical quantities are quadratic forms associated with a sesquilinear
map $q(v,v^*)$ for solution fields $v$ and their complex conjugates $v^*$. 
Examples include the electromagnetic absorption and the electromagnetic energy flux.
For two reasons, the construction of a meaningful modal expansion of sesquilinear forms
$q(v,v^*)$ is not straightforward.
First, the missing orthogonality $q(\tilde{v}_k,\tilde{v}^*_l)\neq 0$ yields
cross terms in the expansion.
Secondly, the conjugation $v^*(\omega_0)$ renders $q(v(\omega_0),v^*(\omega_0))$ nonholomorphic
and the evaluation of this expression for complex eigenfrequencies~$\tilde{\omega}_k$ is problematic.

To derive a modal expansion of sesquilinear quantities with well-defined expansion coefficients,
we extend the framework of the RPE.
The method is based on an analytical continuation
of the sesquilinear form $q(v(\omega_0),v^*(\omega_0))$ from the real axis $\omega_0 \in \mathbb{R}$
into the complex plane $\omega \in \mathbb{C}$. We remark that
$v^*(\omega_0)$ is the solution to ${T}^*({\omega_0}) v^*(\omega_0) = f^*(\omega_0)$.
The system matrix ${T}^*({\omega_0})$ and the right-hand side $f^*({\omega_0})$
have analytical continuations, which we denote by ${T}^\circ({\omega})$ and $f^\circ({\omega})$.
Consequently, the analytical continuation of $v^*({\omega_0})$ reads as
\begin{align}
	v^\circ(\omega) = {T}^\circ({\omega})^{-1} f^\circ(\omega). \label{holo_sesq}
\end{align}
Finally, this gives the analytical continuation $q(v(\omega),v^\circ(\omega))$
into the complex plane and the modal expansion can be computed.

Note that if a solution of the coupled system given by Eqs.~\eqref{coupled_system1}
and \eqref{coupled_system2}  has a pole in $\omega = \tilde{\omega}_k$,
then its complex conjugate has a pole in $\omega = {\tilde{\omega}^*_k}$.
Thus, $q(v(\omega),v^\circ(\omega))$ has poles in $\tilde{\omega}_k$
and also in ${\tilde{\omega}^*_k}$.
This has to be taken into account for the RPE. 
The calculation of a modal quantity corresponding to a specific  $\tilde{\omega}_k$
involves the summation of the Riesz projections for $\tilde{\omega}_k$ and $\tilde{\omega}^*_k$.

As the derivation of $v^\circ(\omega)$ is given formally,
we remark, for a better physical understanding, that
the complex conjugation of the system matrix and the right-hand side
corresponds to solving the coupled system for $\omega = -\omega_0$
with sign-inverted radiation conditions.

\section{Resonances of a nanowire}\label{sec_iii}  

\begin{figure}[]
	{\includegraphics[width=0.45\textwidth]{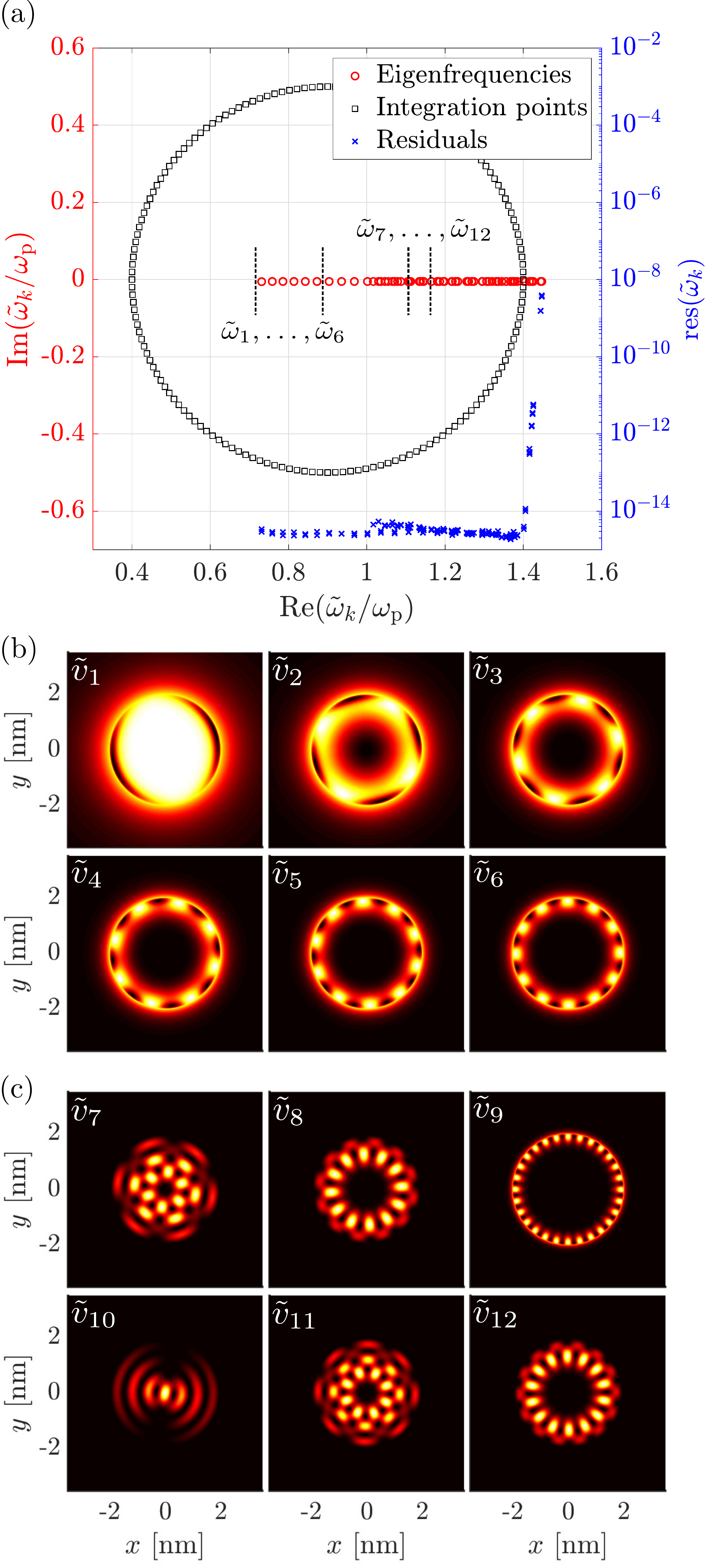}}
	\caption{Eigenfrequencies $\tilde{\omega}_k$ and eigenmodes $\tilde{v}_k$
		of the nanowire.
		(a)~Eigenfrequencies, integration points, and residuals $\mathrm{res}(\tilde{\omega}_k) =
		||T(\tilde{\omega}_k)\tilde{v}_k||_2/||T(\tilde{\omega}_k)||_{F}$, where $||\tilde{v}_k||_2=1$. 
		Inside of the integration contour, $118$ eigenfrequencies are located (including 
		multiplicities).
		(b)~Plots (a.u.) of the electric field intensity of an exemplary
		selection of eigenmodes corresponding to
		eigenfrequencies below the plasma frequency,
		$\tilde{\omega}_1 =	(0.7313 -0.0054i)\omega_\mathrm{p}$,
		$\tilde{\omega}_2 = (0.7585 -0.0050i)\omega_\mathrm{p}$,
		$\tilde{\omega}_3 = (0.7857 -0.0050i)\omega_\mathrm{p}$,
		$\tilde{\omega}_4 = (0.8138 -0.0050i)\omega_\mathrm{p}$,
		$\tilde{\omega}_5 = (0.8429 -0.0050i)\omega_\mathrm{p}$,
		and $\tilde{\omega}_6 = (0.8729 -0.0050i)\omega_\mathrm{p}$.
		(c)~As above, for eigenfrequencies beyond the plasma frequency,
		$\tilde{\omega_7} =	(1.1341 -0.0050i)\omega_\mathrm{p}$,
		$\tilde{\omega}_8 = (1.1373 -0.0050i)\omega_\mathrm{p}$,
		$\tilde{\omega}_9 = (1.1434 -0.0050i)\omega_\mathrm{p}$,
		$\tilde{\omega}_{10} = (1.1453 -0.0050i)\omega_\mathrm{p}$,
		$\tilde{\omega}_{11} = (1.1651 -0.0050i)\omega_\mathrm{p}$,
		and $\tilde{\omega}_{12} = (1.1654 -0.0050i)\omega_\mathrm{p}$.
        Color scale from zero (black) to one (white).}
	\label{fig:fig02}
\end{figure}

We consider a specific setup, a cylindrical metal nanowire which has also been investigated
in the literature, to study HDM-based effects theoretically~\cite{Ruppin_2001}.
For typical nanoplasmonic applications, a quantity of interest is the extinction cross section.
In the following, we first compute eigenfrequencies and eigenmodes of the nanowire.
Based on this, we then investigate the extinction cross section in a modal
sense, i.e., it is shown which of the eigenmodes scatter and absorb an incoming source field and
which of the modes do not interact with the light source.
When the nonlocal HDM is replaced by a local Drude model, only a single resonance is observed
in the extinction cross section~\cite{Ruppin_2001,Schnitzer_ProcRoyal_2016,Hiremath_Hydro2D_2012}.

The investigated sodium nanowire of radius $R = 2~\mathrm{nm}$,
infinitely extended in the $z$ direction [see~Fig.~\ref{fig:fig01}\textcolor{blue}{(a)}], 
is described by 
$\epsilon_\infty = 1$, $\omega_\mathrm{p}=8.65\times10^{15}~\mathrm{s}^{-1}$,
$\gamma = 0.01\,\omega_\mathrm{p}$, and $\epsilon_\mathrm{loc} = \epsilon_0\epsilon_\infty$.
The Fermi velocity is given by $v_\mathrm{F} = 1.07\times10^{6}~\mathrm{m}\mathrm{s}^{-1}$.
The nanowire is surrounded by free space with refractive index equal to one. 
The source field is a  $y$ polarized plane wave  with unit amplitude propagating in the $x$ direction.
For the FEM discretization, a mesh containing about $2000$ triangles
with edge lengths from about $0.05~\mathrm{nm}$ to $1~\mathrm{nm}$ is applied. 
The polynomial degree of the finite elements is set to $p=3$.

The frequency range $0.4\,\omega_\mathrm{p}<\omega_0<1.4\,\omega_\mathrm{p}$
is selected for the modal analysis.
To compute eigenmodes $\tilde{v}_k$ using \textsc{Beyn's Algorithm},
an integration contour around this range is defined.
The parameters for the algorithm are $N = 160$ integration points, $l = 200$
random vectors, and, for the rank drop detection within the SVD,
a tolerance of $\mathrm{tol}_\mathrm{rank} = 10^{-8}$ is chosen.
The SVD and the solution of the resulting small linear
eigenproblem are performed within \textsc{Matlab}.
We obtain $118$ eigenfrequencies inside the integration contour.
The imaginary parts of these eigenfrequencies are 
$\mathrm{Im}(\tilde{\omega}_k) = -0.0050\,\omega_\mathrm{p}$,
except for $\tilde{\omega}_1 =	(0.7313 -0.0054i)\omega_\mathrm{p}$.
We note that the eigenmodes corresponding to eigenfrequencies with
$\mathrm{Im}(\tilde{\omega}_k) = -0.0050\,\omega_\mathrm{p}$ are localized in the nanowire material,
which is modeled with a constant damping $\gamma$.
Other loss channels are not significant for these modes. 
This results in the very similar imaginary parts of the eigenfrequencies.
To numerically assess the quality of the approximations of the eigenfrequencies and
eigenmodes, we compute the residuals
$\mathrm{res}(\tilde{\omega}_k) =||T(\tilde{\omega}_k)\tilde{v}_k||_2/||T(\tilde{\omega}_k)||_{F}$,
where $||\tilde{v}_k||_2=1$. The residuals for eigenfrequencies within the integration contour
are smaller than $6\times10^{-15}$.
The residuals for computed eigenfrequencies
outside the integration contour increase with the distance to the contour.
The integration points, all computed eigenfrequencies, and the residuals
are shown in~Fig.~\ref{fig:fig02}\textcolor{blue}{(a)}.
Plots of the electric field intensity of an exemplary selection of eigenmodes
corresponding to eigenfrequencies in frequency ranges
below and beyond the plasma frequency are shown in~Figs.~\ref{fig:fig02}\textcolor{blue}{(b)}
and~\ref{fig:fig02}\textcolor{blue}{(c)}, respectively.
Note that these eigenfrequencies are semi-simple with an algebraic and geometric multiplicity of two.
The chosen indicies of the eigenfrequencies and eigenmodes are increasing
with increasing real parts of the eigenfrequencies and are
intended to guide the reader through the figures.

Based on the computed spectrum, we investigate the extinction cross section,
\begin{align}
\sigma(\omega_0) = \frac{1}{P_{\mathrm{pw}}}& \left[ \int_{\delta \Omega} \right. 
\frac{1}{2} \mathrm{Re}\left( \mathbf{E}^*(\mathbf{r},\omega_0) \times \mathbf{H}(\mathbf{r},\omega_0)  \right)
\mathrm{d}S   \nonumber \\
&+ \left. \int_{\Omega_\mathrm{nw}} \frac{1}{2}   \mathrm{Re}\left(\mathbf{E}^*(\mathbf{r},\omega_0)
\cdot \mathbf{J}_\mathrm{hd}(\mathbf{r},\omega_0)\right) \mathrm{d}V \right], \nonumber
\end{align}
where the first term is the power flux across the boundary of the entire computational domain,
denoted by $\delta \Omega$,
and the second term is the energy loss in the domain where the nanowire exists,
denoted by $\Omega_\mathrm{nw}$~\cite{Hiremath_Hydro2D_2012}.
The incoming plane wave with real frequencies $\omega_0$ is normalized so that
the power flux through the geometrical cross section of the nanowire is
$P_{\mathrm{pw}} = 4\times10^{-9}~\mathrm{W}$.
To quantify the coupling of the light source to specific eigenmodes, the RPE is applied.
This requires the holomporphic evaluation of sesquilinear quantities from~Eq.~\eqref{holo_sesq}
and yields the modal extinction cross section $\tilde{\sigma}_k(\omega_0)$
corresponding to an eigenfrequency $\tilde{\omega}_k$.
The direct solution of the coupled system given by
Eqs.~\eqref{coupled_system1} and \eqref{coupled_system2}
yields the total extinction cross section $\sigma_\mathrm{tot}(\omega_0)$.

\begin{figure}[]
	{\includegraphics[width=0.45\textwidth]{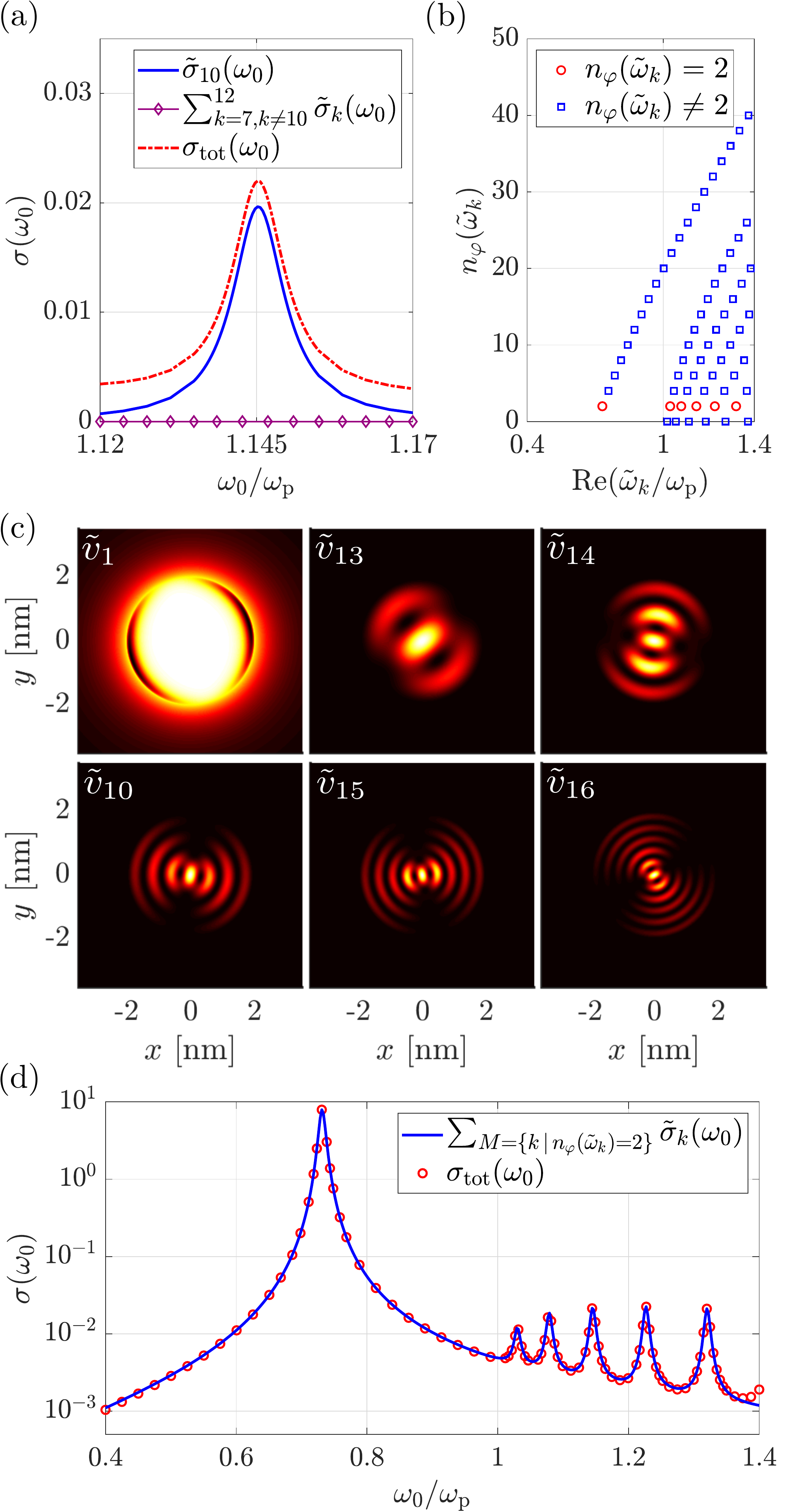}}
	\caption{Modal analysis of
		the extinction cross section $\sigma(\omega_0)$ of the nanowire.
		(a)~$\sigma(\omega_0)$ for the frequency range
		$1.12\,\omega_\mathrm{p}<\omega_0<1.17\,\omega_\mathrm{p}$.
		Modal extinction cross section
		$\tilde{\sigma}_{10}(\omega_0)$ corresponding to the eigenfrequency
		$\tilde{\omega}_{10} = (1.1453 -0.0050i)\omega_\mathrm{p}$ and the sum
		$\sum_{k=7,k\neq 10}^{12} \tilde{\sigma}_k(\omega_0)$
		corresponding to the remaining eigenfrequencies in the frequency range.
        Total extinction cross section $\sigma_\mathrm{tot}(\omega_0)$ for comparison.
		(b)~Classification parameter $n_\varphi(\tilde{\omega}_k)$ for the 
		eigenfrequencies $\tilde{\omega}_k$ in the
		frequency range $0.4\,\omega_\mathrm{p}<\mathrm{Re}(\tilde{\omega}_k)<1.4\,\omega_\mathrm{p}$.
		(c)~Plots (a.u.) of the electric field intensities of the
		eigenmodes with $n_\varphi(\tilde{\omega}_k) = 2$.
        Color scale from zero (black) to one (white).
        (d)~Modal expansion of the extinction cross section 
        $\sum_{M} \tilde{\sigma}_k(\omega_0), M=\{k\,|\,n_\varphi(\tilde{\omega}_k) = 2\}$,
        corresponding to the six eigenfrequencies
		$\tilde{\omega}_1 = (0.7313 -0.0054i)\omega_\mathrm{p}$,
		$\tilde{\omega}_{13} = (1.0301 -0.0050i)\omega_\mathrm{p}$,
		$\tilde{\omega}_{14} = (1.0788 -0.0050i)\omega_\mathrm{p}$,
		$\tilde{\omega}_{10} = (1.1453 -0.0050i)\omega_\mathrm{p}$,
		$\tilde{\omega}_{15} = (1.2267 -0.0050i)\omega_\mathrm{p}$, and
        $\tilde{\omega}_{16} = (1.3202 -0.0050i)\omega_\mathrm{p}$.
        The total extinction cross section $\sigma_\mathrm{tot}(\omega_0)$ is plotted as
        a reference solution. }
	\label{fig:fig03}
\end{figure}

First, we investigate the modal extinction cross section in a small frequency range
including $\tilde{\omega}_{7}, \dots, \tilde{\omega}_{12}$.
Figure~\ref{fig:fig03}\textcolor{blue}{(a)} shows 
$\tilde{\sigma}_7(\omega_0),\dots,\tilde{\sigma}_{12}(\omega_0)$, and $\sigma_\mathrm{tot}(\omega_0)$. 
The eigenmode $\tilde{v}_{10}$ has a significant contribution to $\sigma_\mathrm{tot}(\omega_0)$.
The contributions of the eigenmodes $\tilde{v}_7, \tilde{v}_{8},\tilde{v}_{9},\tilde{v}_{11},$
and $\tilde{v}_{12}$ are negligible.

Secondly, in order to understand why a specific eigenmode couples to the incoming plane wave, a
fast Fourier transform of the electric field intensities of the eigenmodes on a circle inside the
nanowire is performed.
This yields the number of intensity maxima of the eigenmodes along the boundary of the nanowire,
which we denote by $n_\varphi(\tilde{\omega}_k)$.
In this way, it is possible to classify the eigenmodes.
Figure~\ref{fig:fig03}\textcolor{blue}{(b)} shows $n_\varphi(\tilde{\omega}_k)$ for the
frequency range $0.4\,\omega_\mathrm{p}<\mathrm{Re}(\tilde{\omega}_k)<1.4\,\omega_\mathrm{p}$.
The field intensities of the six eigenmodes with  $n_\varphi(\tilde{\omega}_k) = 2$ are 
plotted in Fig.~\ref{fig:fig03}\textcolor{blue}{(c)}.
It can be seen that these modes are dipolelike.
Due to the relation of the radius of the nanowire and the wavelength of the
plane wave, $R\ll\lambda_0$, the overlap integral of source field and eigenmode field has a significant
contribution only for these modes. 

Finally, the modal extinction cross sections $\tilde{\sigma}_k(\omega_0)$ for the 
eigenfrequencies with $n_\varphi(\tilde{\omega}_k) = 2$ are computed.
Figure~\ref{fig:fig03}\textcolor{blue}{(d)} shows the sum of the modal extinction cross sections
$\sum_{M} \tilde{\sigma}_k(\omega_0), M=\{k\,|\,n_\varphi(\tilde{\omega}_k) = 2\}$.
For the investigated scattering of a plane wave, 
the agreement of the expansion with the total extinction cross section $\sigma_\mathrm{tot}(\omega_0)$
demonstrates that the complex scattering behavior of the HDM-based nanowire is governed
by a few eigenmodes only.
Note that the total extinction cross section is in agreement with results from
the literature~\cite{Ruppin_2001,Hiremath_Hydro2D_2012}.

For illumination with different types of source fields, e.g., dipole sources,
also the remaining eigenmodes of the rich spectrum can be excited.

\vspace{-0.15cm}

\section{Conclusions}\label{sec_iv}
We investigated the light-matter interaction in nanoplasmonic systems described by the HDM.
We presented a contour-integral-based framework for modal analysis,
which enables the direct computation of the spectrum of nonlocal material systems. 
We introduced an approach for the modal expansion of sesquilinear quantities.
This opens the possibility to investigate typical physical observables, e.g., 
the energy flux, the energy absorption, and overlap integrals for extraction efficiencies.
Due to the generality of this approach, we expect that it will prove useful also
in other fields of physics. Resonant states and the modal extinction cross section
of a metal nanowire were calculated. While the spectrum of this system consists
of many eigenfrequencies, only a few resonant states
have a significant contribution to the extinction cross section. These resonant
states were identified and used to expand the quantity of interest.

As demonstrated, nanoplasmonic systems on small length scales
exhibit a large number of additional resonant states described by the HDM. 
A typical feature of these states is their high local field energy concentration.
With precisely defined source fields, specific states can be excited. 
We expect that this will allow for additional degrees of freedom in tailoring
light-matter interactions. A modal picture is a prerequisite for the understanding
and for the design of corresponding nanoplasmonic devices. 

\vspace{-0.15cm}

\section*{Acknowledgements}
We acknowledge Philipp-Immanuel Schneider and Fridtjof Betz for fruitful discussions.
We acknowledge funding by the Deutsche Forschungsgemeinschaft  
under Germany's Excellence Strategy - The Berlin Mathematics 
Research Center MATH+ (EXC-2046/1, Project No. 390685689, AA4-6). 
This work is partially funded through the project 17FUN01 (BeCOMe)
within the Programme EMPIR.
The EMPIR initiative is co-founded by the European Union's Horizon 2020
research and innovation program and the EMPIR Participating Countries.

\end{document}